\documentclass[prb,aps, superscriptaddress,twocolumn]{revtex4-1}
\usepackage{amsmath,amssymb}
\usepackage{graphicx}
\usepackage{wasysym}
\usepackage{amsfonts}
\usepackage{bm}
\usepackage{enumerate}
\usepackage{color}
\usepackage[resetlabels]{multibib}
\usepackage{epstopdf}
\usepackage{latexsym}
\usepackage[breaklinks,colorlinks = true,linkcolor = red,urlcolor=cyan,citecolor=red]{hyperref}
\usepackage[caption=false,singlelinecheck=false]{subfig}

\usepackage{times}
\newcommand{\bea}{\begin{eqnarray}}
\newcommand{\eea}{\end{eqnarray}}
\newcommand{\be}{\begin{eqnarray}}
\newcommand{\ee}{\end{eqnarray}}
\newcommand{\bw}{\begin{widetext}}
\newcommand{\ew}{\end{widetext}}

\newcommand\numberthis{\addtocounter{equation}{1}\tag{\theequation}}

\begin{document}
\title{Controllable Floquet edge modes in a multi-frequency driving system}
\author{HaRu K. Park}
\email{haru.k.park@kaist.ac.kr}
\affiliation{Department of Physics, Korea Advanced Institute of Science and Technology, Daejeon, 34141, Korea}
\author{Junmo Jeon}
\email{junmo1996@kaist.ac.kr}
\affiliation{Department of Physics, Korea Advanced Institute of Science and Technology, Daejeon, 34141, Korea}
\author{Gil Young Cho}
\affiliation{Department of Physics, Pohang University of Science and Technology (POSTECH), Pohang 37673, Republic of Korea}
\affiliation{Center for Artificial Low Dimensional Electronic Systems, Institute for Basic Science (IBS), Pohang 37673, Republic of Korea}
\affiliation{Asia Pacific Center for Theoretical Physics, Pohang 37673, Republic of Korea}
\author{SungBin Lee}
\email{sungbin@kaist.ac.kr}
\affiliation{Department of Physics, Korea Advanced Institute of Science and Technology, Daejeon, 34141, Korea}
\date{\today}
\begin{abstract}
A driven quantum system has been recently studied in the context of nonequilibrium phase transitions and their responses. 
In particular, for a periodically driven system, its dynamics are described in terms of the multi-dimensional Floquet lattice with a lattice size depending on number of driving frequencies and their rational or irrational ratio.  
So far, for a multi-frequency driving system, the energy pumping between the sources of frequencies has been widely discussed as a signature of topologically nontrivial Floquet bands. However, the unique edge modes emerging in the Floquet lattice has not been explored yet. 
Here, we discuss how the edge modes in the Floquet lattice are controlled and result in the localization at particular frequencies, when multiple frequencies are present and their magnitudes are commensurate values. 
First, we construct the minimal model to exemplify our argument, focusing on a two-level system with two driving frequencies. For strong frequency limit, one can describe the system as a quasi-one dimensional Floquet lattice where the effective hopping between the neighboring sites depends on the relative magnitudes of potential for two frequency modes. With multiple driving modes, there always exist the non-trivial Floquet lattice boundaries via controlling the frequencies and 
this gives rise to the states that are mostly localized at such Floquet lattice boundaries, i.e. particular frequencies. We suggest  the time-dependent Creutz ladder model as a realization of our theoretical Hamiltonian and show the emergence of controllable Floquet edge modes.   
\end{abstract}


\maketitle

\section{Introduction}
Topology becomes an essential concept in modern condensed matter physics \cite{PhysRevLett.95.226801, PhysRevLett.95.146802, bernevig2006quantum, hasan2010colloquium,qi2011topological,xiao2020non,helbig2020generalized}. One such example is the so-called Thouless pumping model, which is an adiabatic, time-dependent model. The model is designed to pump an integer number of electric charge, which is related to the topological winding number\cite{PhysRevB.27.6083, switkes1999adiabatic,PhysRevLett.109.106402, lohse2016thouless, nakajima2016topological}. More recently, it has been noted that non-trivial topological properties can also emerge in the time-periodic driving systems\cite{PhysRevB.82.235114,PhysRevX.3.031005,PhysRevX.7.041008,rudner2020band}. Such time-periodic models are called as the Floquet model, which has been extensively investigated in the context of transports, laser controlled atoms and electron-phonon coupled systems\cite{PhysRev.138.B979,PhysRevA.7.2203,PhysRevB.66.205320,GRIFONI1998229,chu2004beyond,murakami2017nonequilibrium, PhysRevX.5.031001,PhysRevB.90.195429,PhysRevX.4.031027,RevModPhys.89.011004}. It has been also noted that a $d$-spatial-dimensional time-dependent system under the $D$-frequency drives 
can be classified by the topology of the static Hamiltonian in $(d+D)$ spatial dimensions and the relevant examples are studied.\cite{PhysRevLett.126.106805,PhysRevB.104.224301,PhysRevLett.127.166804}

In this paper, we will uncover another novel phenomena of the driven quantum systems. Specifically, we will show that the multi-frequency driving system  can induce the localized ``edge" mode in the frequency lattice, equivalently the Floquet lattice (which will be defined below). These novel topological modes are localized at a special frequency. The physical origin of such novel mode can be understood as follows. When the ratio between the frequencies are commensurate to each other, one can consider the multi-dimesional Floquet lattice, which repeats along a certain direction as shown in Fig.\ref{fig:floquet-lattice}. This naturally introduces an ``edge" to the Floquet lattice, and the edge may trap an interesting mode, which depends on the topology of the driven system. As a proof of this claim, we will introduce an explicit, two-level model with two commensurate driving frequency, which can explicitly demonstrate the desired physics. 

The rest of this paper is organized as follows. First, we briefly review the Floquet theory in a periodically driven system. Then we introduce our model, a two-level system with two driving modes, which is transformed into two-dimensional Floquet lattice. In the strong frequency limit, the Floquet lattice can be mapped onto a quasi-one dimensional lattice having a non-trivial edge mode. We also discuss the possible realization of our model in the quasi-one dimensional Creutz lattice. Finally, we also support our theoretical argument with various numerical results.

\section{Floquet Lattice \& Floquet Theory}
Here, we briefly review the basic physics of the Floquet theory,\cite{Floquet1883} describing a system under the multiple time-periodic driving. Our starting point is the Hamiltonian $H$ which depends on the $D$-time periodic parameters $(\theta_1,\cdots,\theta_D)=\vec{\theta}_D$, each depends on one time parameter $t$. That is, with some constants $(\Omega_1,\cdots,\Omega_D)=\vec{\Omega}$,
\begin{equation}\label{eqn:Hamiltonian}
  H(\theta_i)=H(\theta_i+2\pi),\quad \theta_i(t)=\Omega_i t.
\end{equation}
We attempt to solve the time-dependent Schr{\"o}dinger equation, 
\begin{align}
i\partial_t |\psi(t)\rangle = H(\vec{\theta}(t))|\psi(t)\rangle. 
\end{align} 
Performing the Fourier transformation, we obtain, 
\begin{equation}\label{eqn:fft-schrodinger}
  \omega|\psi(\omega)\rangle = \sum_{\vec{m}} H_{\vec{m}}|\psi(\omega-\vec{m}\cdot \vec{\Omega})\rangle.
\end{equation}
Here, $|\psi(\omega)\rangle$ and $H_{\vec{m}}$ are the Fourier coefficients of the wavefunction and the Hamiltonian respectively.
\begin{equation}
  |\psi(t)\rangle=\int d\omega e^{-i\omega t}|\psi(\omega)\rangle,\quad H(\vec{\theta})=\sum_{\vec{m}}e^{-i\vec{m}\cdot \vec{\theta}} H_{\vec{m}}.
\end{equation}
Note that Eq.\eqref{eqn:fft-schrodinger} only couples the frequency $\epsilon$ with other frequencies $\omega=\epsilon+\vec{n}\cdot \vec{\Omega}$. Therefore, by indexing $|\vec{n}\rangle \equiv |\psi(\epsilon+\vec{n}\cdot\vec{\Omega})\rangle,$ Eq.\eqref{eqn:fft-schrodinger} becomes,
\begin{equation}\label{eqn:floquet-hopping}
  \epsilon|\vec{n}\rangle = \sum_{\vec{m}} (H_{\vec{n}-\vec{m}}-\vec{n}\cdot \vec{\Omega}\delta_{\vec{n},\vec{m}})|\vec{m}\rangle.
\end{equation}
Eq.\eqref{eqn:floquet-hopping} is mathematically equivalent with the $D$-dimensional tight-binding model defined on a lattice at sites $\vec{n}$. We call this lattice as a ``Floquet lattice". In this analogy, $|\vec{n}\rangle$ represents the state, when a particle is exactly at the site $\vec{n}$ on the lattice. We will also call $\epsilon$ as the quasi-energy. Physically, one can interpret the numbers $n_i$ (the $i$-th component of the vector $\vec{n}$) as the number of absorbed photons from the $i$-th drive. Therefore, the hopping terms, $H_{\vec{n}}$ in Eq.\eqref{eqn:floquet-hopping}, describe the process of absorbing and/or emitting certain number of photons. 
Unlike an ordinary tight-binding model, Eq.\eqref{eqn:floquet-hopping} has an additional on-site potential $-\vec{n}\cdot \vec{\Omega}$. This on-site potential term corresponds to the electric field $\vec{\Omega}$ on the tight-binding model, hence we call $\vec{\Omega}$ as the quasi-electric field.

\section{The Model}
Let us consider a two-frequency-driven Hamiltonian on a two-level system in the rest of this paper. The Hamiltonian is represented as, 
\begin{align*}\label{eq:main-hamiltonian}
  H(t)=B_z(k)\sigma_z &+ \left[(\Delta-\delta(k))\cos\left(p\Omega t\right)\right.\\
  &\left. +(\Delta+\delta(k))\cos\left(q\Omega t\right)\right]\sigma_x.\numberthis
\end{align*}
Here, $k$ is another parameter (periodic in $2\pi$) which can be independently tuned. This variable will be used to change the parameters $B_z$ and $\delta$, which will be shown later to play the role of staggered potential and alternating hopping in the Floquet lattice. $\sigma_x$ and $\sigma_z$ are the Pauli matrices, whose basis can be considered as a pseudo-spin states $\{|\uparrow\rangle, |\downarrow\rangle\}$. $t$ represents the time and $\Omega$ is the frequency of the system with coprime integers $p$ and $q$.

Eq.\eqref{eq:main-hamiltonian} describes a time-periodic system with the period $T=2\pi/ \Omega$. Hence, it can be Fourier transformed into the discrete frequency domain. After performing the Fourier transformation on time, we can write out the following tight-binding model on the Floquet lattice, which is equivalent to the Schr{\"o}dinger equation on the following Hamiltonian $H_F$.
\begin{align*}\label{eq:floquet-hamiltonian}
  H_{F}=\sum_{\vec{n},\alpha,\beta}&\left[B_z(k)\sigma_z^{\alpha\beta}-\vec{n}\cdot\vec{\Omega}\delta^{\alpha\beta}\right]c_{\vec{n};\alpha}^\dagger c_{\vec{n};\beta}\\
  &+(\Delta-\delta(k))\sigma_x^{\alpha\beta}c_{\vec{n}+(1,0);\alpha}^\dagger c_{\vec{n};\beta}+h.c.\\
  &+(\Delta+\delta(k))\sigma_x^{\alpha\beta}c_{\vec{n}+(0,1);\alpha}^\dagger c_{\vec{n};\beta}+h.c.\numberthis 
\end{align*}
Here, $c_{\vec{n};\alpha}$ represents the annihilation operator for a pseudo-spin $\alpha$ state with frequency $\epsilon+\vec{n}\cdot \vec{\Omega}$, where $\epsilon$ is the quasi-energy. Here $\vec{\Omega} = \Omega(p, q)$. 

Our primary goal here is to show that the Hamiltonian Eq.\eqref{eq:floquet-hamiltonian} can support the topological ``edge" modes, which are localized in the corresponding Floquet lattice. We will demonstrate this in a few different ways. 

\begin{figure}[t]
  \subfloat[]{\label{fig:lattice-structure}\includegraphics[width=0.28\textwidth]{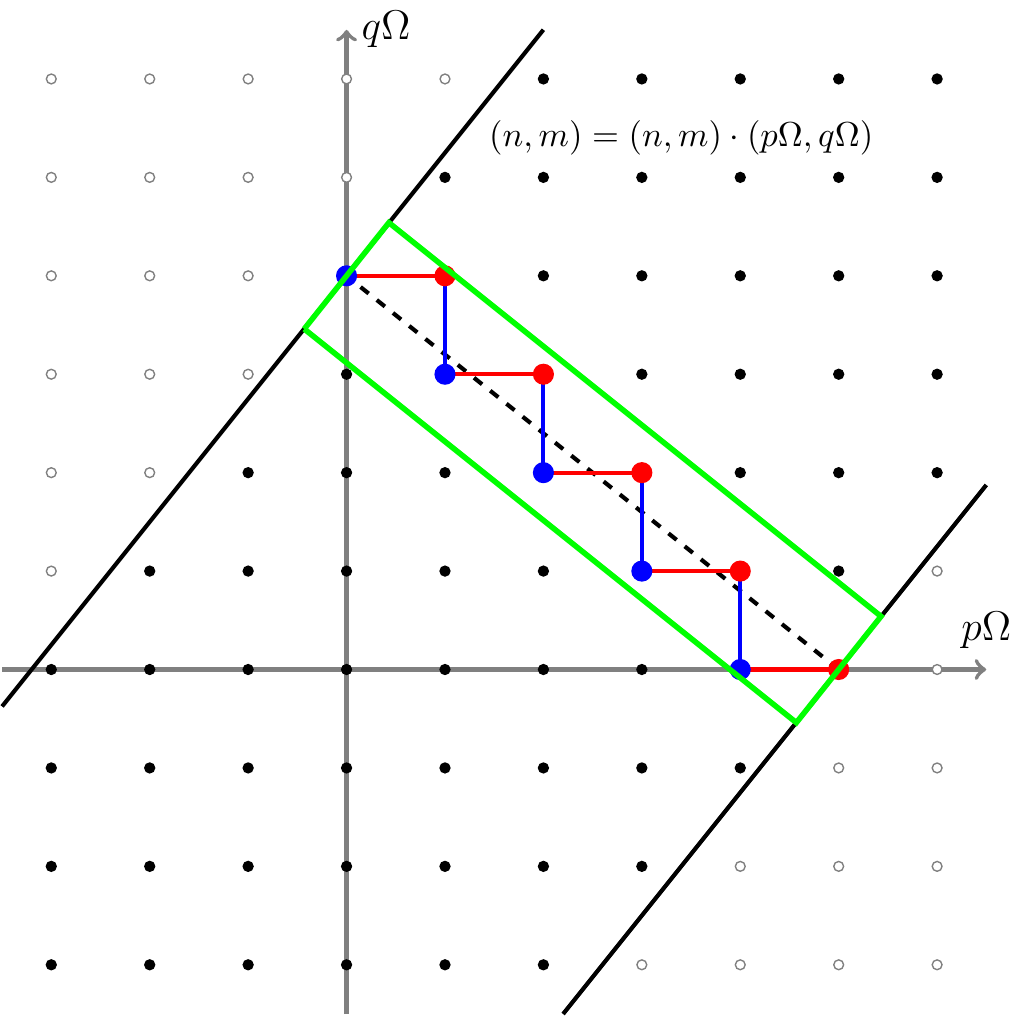}}
  \subfloat[]{\label{fig:lattice-cylinder-structure}\includegraphics[width=0.2\textwidth]{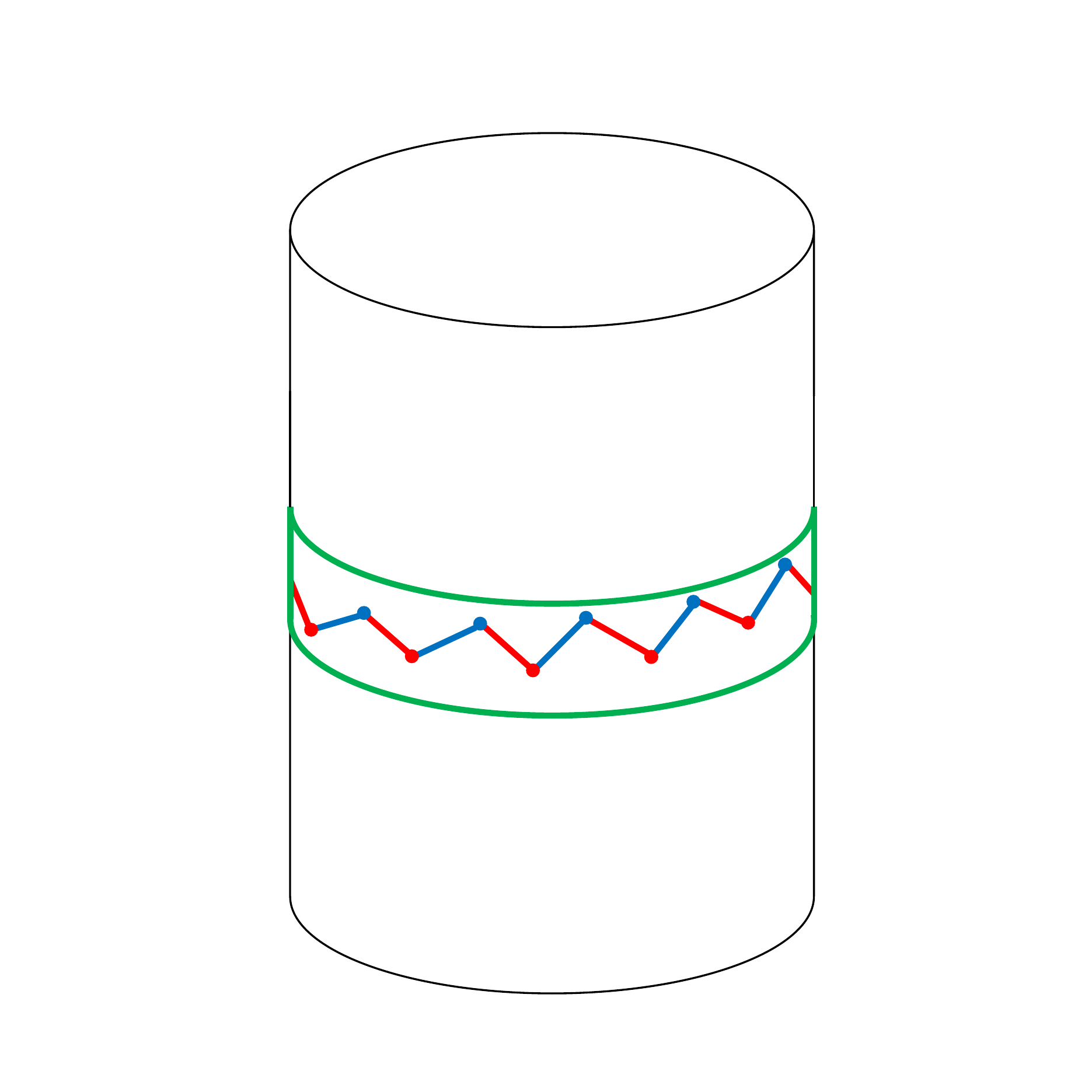}}
  \caption{\ref{sub@fig:lattice-structure} Floquet lattice from Eq.\eqref{eq:floquet-hamiltonian}, with $p=4$ and $q=5$. Each dots represents $|n,m\rangle$ state of the Floquet system, which represents the frequency $(n,m)\cdot(p\Omega, q\Omega)$. Two thick lines are the same line due to the translational symmetry $(n,m)\mapsto (n-q, m+p)$. Dashed diagonal line is an equipotential line, and the red and blue dots in the green box are the points near equipotential line, representing up and down spins respectively. Due to the $\sigma_z$-term, the on-site energy on red and blue dots has different sign, $\pm B_z(k)$. Red horizontal lines and blue vertical lines represents the hopping terms, whose strengths are $\Delta + \delta(k)$ and $\Delta-\delta(k)$, respectively. \ref{sub@fig:lattice-cylinder-structure} By pasting two thick lines in \ref{sub@fig:lattice-structure}, we get the cylindrical structure.}
  \label{fig:floquet-lattice}
\end{figure}

\subsection{Instructive Limit}
The non-trivial topology of Eq.\eqref{eq:floquet-hamiltonian} can be most cleanly manifested in the strong driving limit, i.e., $\Delta\gg B_z(k),  \delta(k)$, so the dominant term is the hopping term with $\sigma_x$. This implies that the distribution of the pseudo-spin is alternating as in Fig.\ref{fig:lattice-structure}. Furthermore, the strong quasi-electric field $\vec{\Omega}$ localizes the ground state around an equipotential line by the Stark localization\cite{RevModPhys.34.645, long2021many}. Thus, in this limit, one can choose the sites near the equipotential line and interactions between those sites to construct an effective quasi-one dimensional lattice. Although there are many equipotential lines in the system, in strong frequency limit, the contribution near the zero frequency mode is dominant. (See Appendix \ref{appendix:stark-localization} for details.)

To exemplify, we first focus on our interest to $q=p+1$. In this case, the sites (in the Floquet lattice) near the equipotential line interacts with the sites connected by the alternating horizontal and vertical bonds, as shown in Fig.\ref{fig:lattice-structure}. Since we can tune the strength of the vertical and horizontal hoppings independently, the quasi-one dimensional lattice becomes an 1D lattice model with a unit-cell which consists of the two sites. In this lattice, the pseudo-spin directions of the two sites are opposite because of $\Delta\gg B_z(k),  \delta(k)$. 

Having these in mind, the total Hamiltonian in Eq.\eqref{eq:main-hamiltonian} can be approximated into the following form.
\begin{align*}\label{eq:floquet-rice-mele}
  H_{1d}(k)=&\sum_n \left[ B_z(k)(c_{2n}^\dagger c_{2n}-c_{2n+1}^\dagger c_{2n+1})\right.\\
  &+(\Delta-\delta(k))c_{2n}^\dagger c_{2n+1}+h.c.\\
  &\left.+(\Delta + \delta(k))c_{2n+1}^\dagger c_{2n+2}+h.c.\right]\numberthis
\end{align*}
Here, $n$ increases along the sites on the Floquet lattice near the equipotential line, following this effective 1D lattice structure in Fig.\ref{fig:lattice-structure}. Here, the parameter $B_z$ plays the role of staggered potential and $\delta$ tunes the alternating hopping strength. 

The topological property of this model in the Floquet lattice can be easily understood by comparing it with the Rice-Mele model\cite{PhysRevLett.49.1455}, which is a time-dependent one-dimensional model with the staggered potential and alternating hopping strength. In this point of view, the Hamiltonian is topologically non-trivial, when the vector $(B_z,\delta)(k)$ winds around the origin of the parameter space. Thus, for example, if we set
\begin{equation}
  B_z(k)=\cos k, \quad \delta(k)=\sin k,
\end{equation}
then the topology of the Hamiltonian becomes nontrivial. More detailed explanation about the Rice-Mele model is described in Appendix \ref{appendix:rice-mele-model}.



\begin{figure}[t]
  \includegraphics[width=0.45\textwidth]{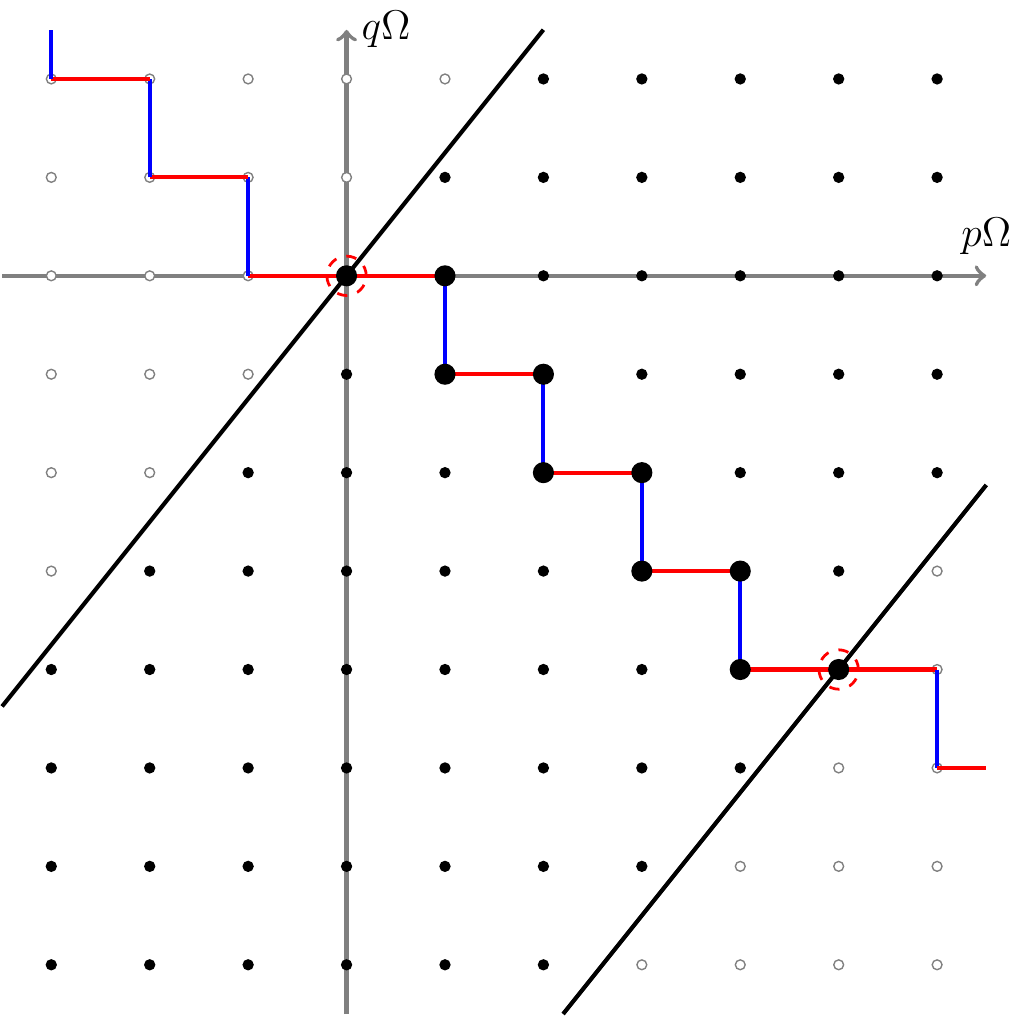}
  \caption{The periodic structure of a quasi-one dimensional Floquet lattice, with $p=4$ and $q=5$. The large black dots represent the points near equipotential line, and horizontal and vertical red and blue lines represent the interaction. Due to the periodicity, the quasi-one dimensional lattice on large black dots goes periodically. Then naturally, the black dots emphasized by red circles become a boundary between two different SSH Hamiltonians.}
  \label{fig:boundary}
\end{figure}

\subsection{Boundary Localization}
We now present the intuitive understanding of the edge mode in the Floquet lattice. Our Floquet Hamiltonian in Eq.\eqref{eq:floquet-hamiltonian} is a periodic Hamiltonian under translation $\vec{n}\rightarrow \vec{n}+(-q,p)$ and has a cylinderical structure, as we can see in Fig.\ref{fig:lattice-cylinder-structure}. This periodicity follows from the fact that $\vec{n}\cdot\vec{\Omega}=0$ when $\vec{n}=(-q,p)$, that is, the system is periodic under the direction perpendicular to the quasi-electric field $\vec{\Omega}$. (Note that $\vec{\Omega} = \Omega (p,q)$.) Therefore, restricting the whole system into the perpendicular direction of $\vec{\Omega}$ must also give a periodic lattice system. However, because the alternating hopping on the quasi-one dimensional lattice starts and ends with the same type of hopping, there exists a boundary between two topologically distinct lattices.

For the limit we consider (See Fig.\ref{fig:lattice-structure}), the 1D lattice has an alternating distribution of hopping term, with two different hoppings $\Delta\pm \delta$. Fig.\ref{fig:boundary} shows the periodic structure of the quasi-1D lattice. Here, the 1D lattice ends and restarts both by the hopping strength $\Delta-\delta$. This introduces a ``solitonic configurion" in the hopping terms, and generates a topological boundary on the periodic lattice. This topological boundary exists due to the one-directional periodicity of the lattice, and thus only happens when the ratio between frequencies are commensurate.


To concretely demonstrate the edge mode, we consider the easiest case $B_z(k)=\cos k$ and $\delta(k)=\sin k$, and compare $k=\pi/2$ and $k=3\pi/2$ case. In both cases, we have $B_z(k)=0$. Hence the 1D lattice becomes a Su-Schreiff-Heeger(SSH) lattice with hopping terms $\Delta\pm 1$, and the boundary of the system can be considered as the joint between the trivial and topological phases of the SSH chain. This creates the localized eigenstate near the boundary. Because each site of the Floquet lattice represents the frequency of the state, this corresponds to the state with a high occupation on the frequency which represents the boundary. Since the small perturbation does not affects much to the localization of the state, the localization appears on every $k\in [0,2\pi]$ for $\delta(k) \neq 0$. 

Until now we have only discussed $q=p+1$ case. In this case, the position of the localized edge mode is not controllable, since there is only one boundary point and the state is simply localized near it. To show the controllability of the localized mode, we introduce $q=p+2$ case, which can be considered as the case with multiple boundaries.

\begin{figure}[t]
  \includegraphics[width=0.45\textwidth]{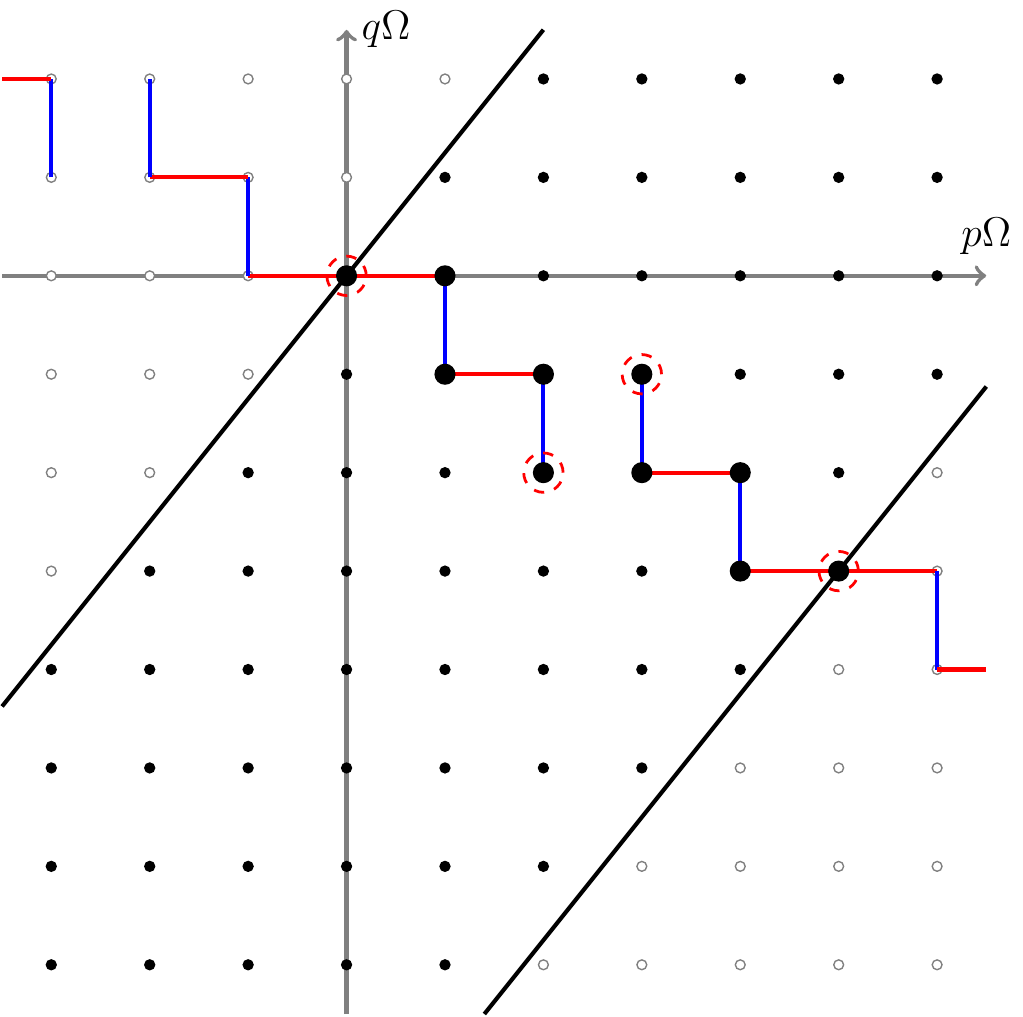}
  \caption{The periodic structure of a quasi-one dimensional Floquet lattice, with $p=3$ and $q=5$. The large black dots represent the points near equipotential line, and horizontal and vertical red and blue lines represent the interaction. The red dashed circles represent the position of boundary on quasi-one dimensional lattice.}
  \label{fig:boundary_2}
\end{figure}

Fig.\ref{fig:boundary_2} shows an example of quasi-one dimensional lattice structure for $q=p+2$, particularly when $p=3$ and $q=5$, i.e. $q=p+2$. In the vicinity of the equipotential line, there are in total three different boundary points, marked as dotted circles in Fig.\ref{fig:boundary_2}. Specifically, when the strength of the blue bond is stronger than the red one, we expect the localization at $(0,0)$ point, which gives the same result with $q=p+1$ case. On the other hand, when the strength of the red bond is stronger then the blue one, the localization in the middle of the quasi-one dimensional lattice would occur. Since the strengths of red and blue bonds depend on the parameter $k$, we can control the localization mode of the system by tuning $k$ value.



\subsection{Physical Realization via Creutz Lattice}
Before showing some numerical results which justify our heuristic understandings above, we present a potential realization of our model in a lattice model. So far, we have not assigned any particular physical meaning to $k$, and hence, in principle, there could be many different ways to realize the Hamiltonian in Eq.\eqref{eq:main-hamiltonian}. Here, we will consider a time-dependent 1D lattice, which is a variation of the so-called Creutz ladder\cite{PhysRevLett.83.2636}, as a potential candidate to realize our model. 

To realize the Hamiltonian in Eq.\eqref{eq:main-hamiltonian}, we consider $k$ as the momentum parameter. Furthermore, by considering two sublattices $A,B$ in real space, we can choose the basis $\{|\uparrow\rangle, |\downarrow\rangle\}$ of $\sigma_x, \sigma_z$. We will assign the creation operator of the momentum $k$ at $A,B$ sites in a way that $|\uparrow\rangle_{k}=c_{k,A}^\dagger|0\rangle$ and $|\downarrow\rangle_{k}=c_{k,B}^\dagger|0\rangle$. This allows us to write, 
\begin{equation}
   \sigma_x\rightarrow c_{k,A}^\dagger c_{k,B}+c_{k,B}^\dagger c_{k,A},~~~~ \sigma_z\rightarrow c_{k,A}^\dagger c_{k,A}-c_{k,B}^\dagger c_{k,B}.
\end{equation}
We can now write down the Hamiltonian with $B_z(k)=B_z\cos k$ and $\delta(k)=\delta \sin k $ as following.
\begin{align*}
  &H(k, t)=B_z \cos k \left(c_{k,A}^\dagger c_{k,A}-c_{k,B}^\dagger c_{k,B}\right)\\
  &+\Delta (\cos(p\Omega t)+\cos(q\Omega t))\left( c_{k,A}^\dagger c_{k,B} + c_{k,B}^\dagger c_{k,A}\right)\\
  &+\delta \sin k (-\cos(p\Omega t)+\cos(q\Omega t))\left( c_{k,A}^\dagger c_{k,B} + c_{k,B}^\dagger c_{k,A}\right)\numberthis.
\end{align*}
Using the inverse Fourier transformation, $c_{A/B,k} = \sum_x c_{A/B, x} e^{-ikx}$,
one gets,
\begin{equation}\label{eq:creutz-ladder}
  H(t)\equiv \sum_k H(k,t) = H_h + H_v + H_d.
\end{equation}
Here, the horizontal interaction between dimers of the ladder, $H_h$, the intra-interaction of the dimers, $H_v$, and the diagonal interaction between dimers of the ladder, $H_d$, are represented as respectively,
\begin{align*}
  H_h&= \frac{B_z}{2} \sum_x \left[c_{x+1,A}^{\dagger}c_{x,A}+ c_{x+1,B}^\dagger c_{x,B}+h.c.\right], \\
  H_v&= \frac{\Delta}{2} (\cos(p\Omega t)+\cos(q\Omega t))\sum_x \left[c_{x,A}^{\dagger}c_{x,B}+h.c.\right], \\
  H_d&=\frac{1}{4}(\!-\!\cos(p\Omega t) \!+\! \cos(q\Omega t))\sum_x \Big[ic_{x+1,A}^\dagger c_{x,B} \\
   & ~~~~~~~ +ic_{x+1,B}^\dagger c_{x,A}+h.c.\Big]. \numberthis
\end{align*}
 This type of ladder is so called the Creutz ladder. After constructing time-dependent Creutz ladder with coefficients given above, we may scan the system on the momentum space $k$. This process experimentally measures the localization of the state, which is the clue of the edge states in our model.

\section{Numerical Demonstrations} 
In this section, we will present the numerical proof of our claim above. In this numerical simulation, we solve the full Floquet problem and do not restrict ourselves to the effective 1D Floquet Hamiltonian. We will find that the above understanding based on the effective 1D lattice model in Eq.\eqref{eq:floquet-rice-mele} is indeed correct. 

We first confirm the localization of the modes under the strong-enough quasi-electric field $\vec{\Omega}$, or equivalently, on the strong frequency limit. Numerically, we show the localization near equipotential line when the energy level of the frequency $\Omega$ is comparable with the hopping term $\Delta$. This is reasonable because if the quasi-electric field is weaker than the hopping term, then the hopping term leads the state spread over all the Floquet lattice. Detailed numerical studies are explained in Appendix \ref{appendix:stark-localization}.

\begin{figure}[t]
  \subfloat[]{\label{fig:localization-k0}\includegraphics[width=0.24\textwidth]{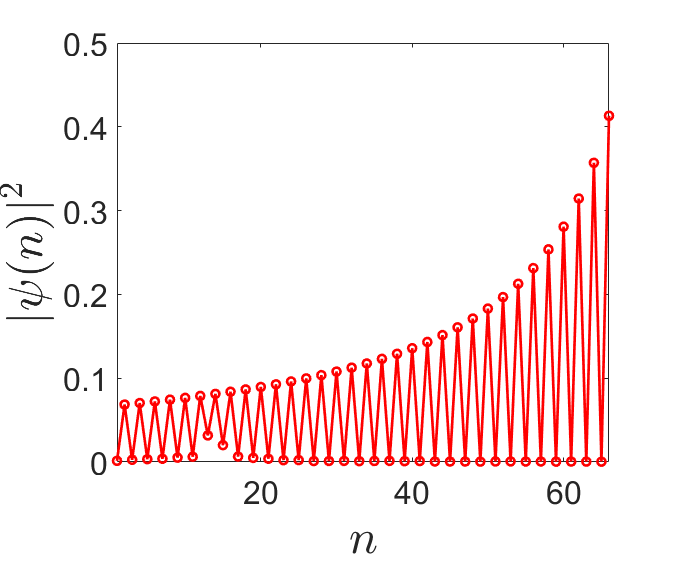}}
  \subfloat[]{\label{fig:localization-kpi}\includegraphics[width=0.24\textwidth]{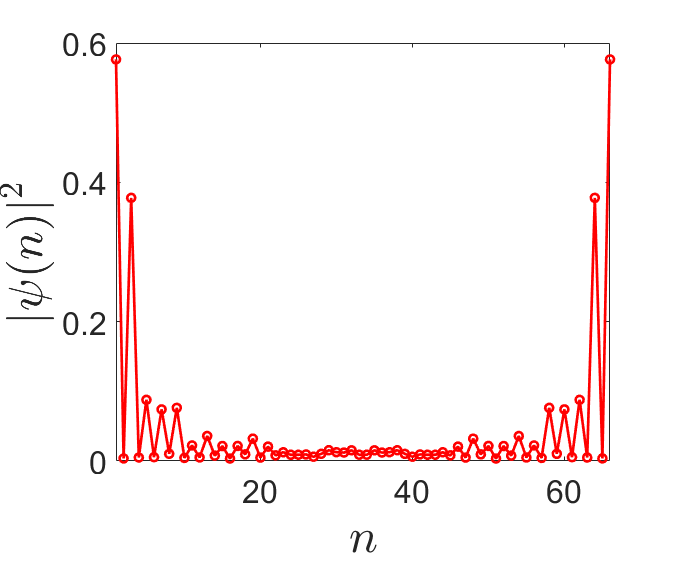}}
  \caption{Distribution of the state on the equipotential line of the Floquet lattice at \ref{sub@fig:localization-k0} $k=0$ and \ref{sub@fig:localization-kpi} $k=\pi/2$. Here, the calculation is performed with $\Delta=10, \Omega=10, \delta(k)=\sin k, B_z(k)=\cos k$, and $p=33, q=34.$}
  \label{fig:localization-numerical}
\end{figure}

Now let us explore the physics of the topological edge mode. We plot the distribution of the states on the quasi-one dimensional Floquet lattice, at $k=0$ (Fig.\ref{fig:localization-k0}) and $k=\pi/2$ (Fig.\ref{fig:localization-kpi}) respectively. 
For $k=0$, i.e., $\delta=0$, it exhibits the extended state in addition to the localized mode at zero frequency. Whereas, for $k=\pi/2$ i.e., $\delta \neq 0 $, the localized state at zero frequency only survives, which can be easily understood from the SSH model as discussed above. 
Fig.\ref{fig:localization-numerical} shows the distribution of the states projected on the alternating pseudo-spin state, for example, $|\uparrow\downarrow\uparrow\downarrow\cdots\rangle$. Suppose that we projected the state on another alternating pseudo-spin configuration, $|\downarrow\uparrow\downarrow\uparrow\cdots\rangle$. Because this process changes up state and down state, we may consider the total Hamiltonian transforms as $\sigma_z\rightarrow -\sigma_z$. 
By absorbing the sign change of $\sigma_z$ into $B_z(k)=B_z\cos(k)$ parameter, we may consider this transformation as  following.
\begin{align*}
  B_z\cos(k)&\rightarrow -B_z\cos(k)=B_z\cos(\pi-k),\\
  \delta\sin(k)&\rightarrow \delta\sin(k)=\delta\sin(\pi-k)\numberthis
\end{align*}
Therefore, flipping the sign is equivalent to transforming $k$ into $\pi-k$. This shows that $k=0$ case in Fig.\ref{fig:localization-k0} and $k=\pi$ case changes in the different alternating pseudo-spin projection case. Notice that this also implies the opposite sign of winding numbers of $(B_z, \delta)$ and $(-B_z, \delta)$, showing the boundary is created by the joint between two lattices with different topological properties.

To see the controllable edge mode, we discuss the case with $q=p+2$ with $p,q$ both odd, which also have a quasi-one dimensional lattice near the equipotential line with different structure.

\begin{figure}[t]
  \subfloat[]{\label{fig:localization-k0-oddodd}\includegraphics[width=0.24\textwidth]{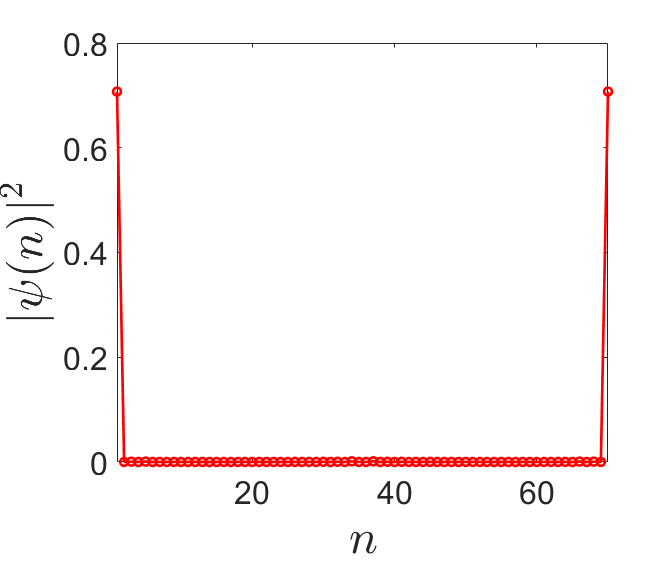}}
  \subfloat[]{\label{fig:localization-kpi-oddodd}\includegraphics[width=0.24\textwidth]{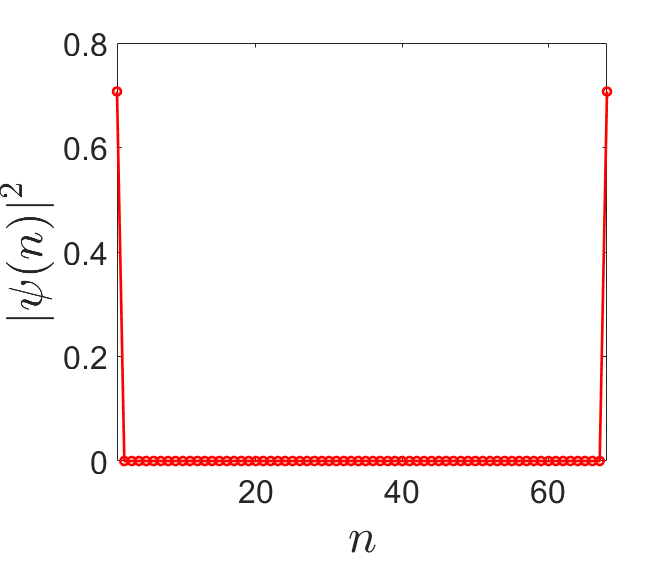}}
  
    \subfloat[]{\label{fig:localization-k0eps-oddodd}\includegraphics[width=0.24\textwidth]{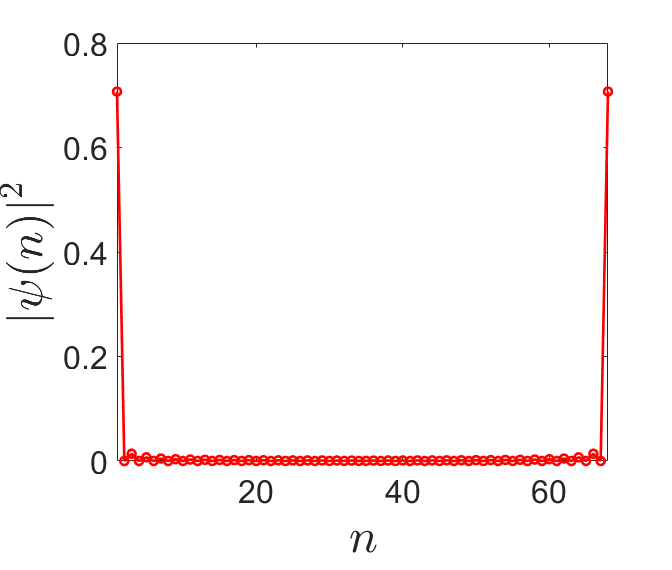}}
  \subfloat[]{\label{fig:localization-kpieps-oddodd}\includegraphics[width=0.24\textwidth]{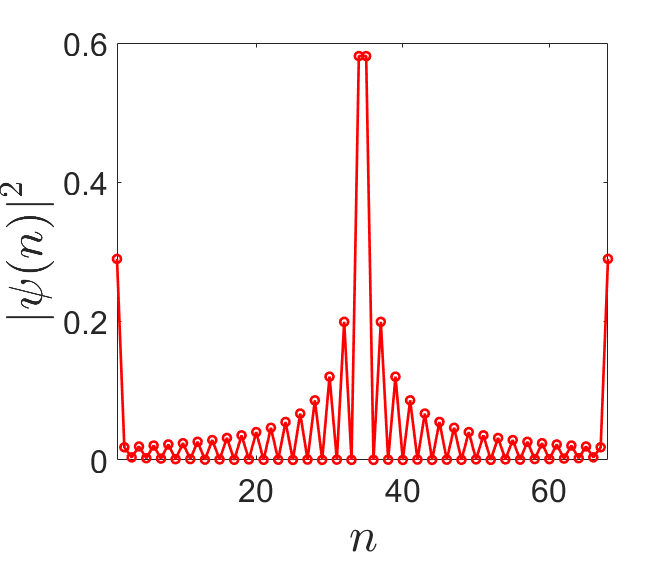}}
  \caption{Distribution of the state on the equipotential line of the Floquet lattice at \ref{sub@fig:localization-k0-oddodd} $k=\pi/2$, \ref{sub@fig:localization-kpi-oddodd} $k=3\pi/2$, \ref{sub@fig:localization-k0eps-oddodd} $k=\pi/2+\eta$, and \ref{sub@fig:localization-kpieps-oddodd} $k=3\pi/2+\eta$. Here, $\Delta=10, \Omega=10, \delta(k)=\sin(k), B_z(k)=\cos(k), \eta=0.001, $ and $p=33, q=35$.}
  \label{fig:localization-numerical-oddodd}
\end{figure}

Figs.\ref{fig:localization-k0-oddodd} and \ref{fig:localization-kpi-oddodd} represent the exact probability distributions on the quasi-one dimensional lattice, for $k=\pi/2$ and $k=3\pi/2$ respectively. Theoretically, we have expected the localized mode at the edge of quasi-one dimensional lattice when $k=\pi/2$, and at the center when $k=3\pi/2$. However, numerical result does not show the localization at the center for both $k=\pi/2$ and $3\pi/2$. Instead, if we measure on $k=3\pi/2+\eta$ with small $\eta$ value, then we get the localization behavior at the center as desired as shown in Fig.\ref{fig:localization-kpieps-oddodd}.
The absence of the localized mode at the center for $k=3\pi/2$ is originated from the on-site energy induced by the quasi-electric field. Ignoring the on-site energy induced by the quasi-electric field, the states on quasi-one dimensional lattice are three-fold degenerate excluding the spin degrees of freedom: two of them are localized at the center of the quasi-one dimensional lattice, and one of them are localized at the edge. Because the edge of the quasi-one dimensional lattice is closer to the equipotential line, the state localized at the edge is more stable compared to the other ones when $k=3\pi/2$. Thus, we observe the localization dominant at the edge, as shown in Fig.\ref{fig:localization-kpi-oddodd}.

However, as the value $k$ deviates from $3\pi/2$, the degeneracy ignoring quasi-electric field breaks down, and for non-zero quasi energy, one can observe the states localized at the center. Thus, when $k$ sufficiently deviates from $3\pi/2$, the localized state at the center is shown for non-zero quasi energy despite the presence of the quasi-electric field (See Fig.\ref{fig:localization-kpieps-oddodd}.). Notice that because only a single boundary exist at the edge for $k=\pi/2$ case, the localization at the center does not appear although we slightly change the $k$-value from $\pi/2$, as we can see in \ref{fig:localization-k0eps-oddodd}.

In summary, we have shown that the localization modes at particular frequencies can be controlled via multi-frequency ratio and their magnitudes, and this can be explained by the transformation of our model into the Floquet SSH model on a quasi-one dimensional lattice.

\section{Conclusions}
In this research, we designed a one dimensional Creutz ladder model with two-driving modes. Under strong frequency, the Floquet version of this model can be reduced into quasi-one dimensional model with nontrivial topological properties. Due to the construction, this quasi-one dimensional model  is mathematically equivalent to the time-driven SSH model with boundary. Therefore the localization on the Floquet lattice becomes the evidence of the topological property, generated by the junction of two SSH models with different topological properties, which is experimentally measurable. 

In multi-frequency system with their commensurate frequency ratio, our work suggested a  new method to build the boundary on Floquet lattice. This localization can be measured experimentally, when the frequency scale is comparable to the amplitude of driving modes, by investigating the frequency profile of the eigenstate. 
Because we have shown the topological property which occurs due to the interplay between spatial dimension and frequency  modes, the extension of our system can be done not only by adding driving modes but also increasing the spatial dimension. This makes the possibility of the topological system on spatial multi-dimension with multi-frequency drives, suggesting new kinds of topological Floquet systems.


\begin{acknowledgments}

\noindent	
{\em Acknowledgments.---}
This work is supported by the National Research Foundation Grants (NRF- 2020R1A4A3079707, NRF grand 2021R1A2C1093060). GYC acknowledges the support of the National Research Foundation of Korea (NRF) funded by the Korean Government No. 2020R1C1C1006048 and the support by IBS-R014-D1. This work is also supported by the Air Force Office of Scientific Research under Award No. FA2386-20-1-4029. GYC acknowledges financial support from Samsung Science and Technology Foundation under Project Number SSTF-BA2002-05.
\end{acknowledgments}

\appendix
\section{Rice-Mele model}\label{appendix:rice-mele-model}
The Rice-Mele model is a time-periodic model with adiabatic charge pumping, with non-trivial topological property. As a one-dimensional lattice with a unit cell containing two sublattices, its hamiltonian can be written as following.
\begin{align*}\label{eqn:rm-model}
  H(t)=&\sum_n \left[\delta h(t)(c_{n,1}^\dagger c_{n,1}-c_{n,2}^\dagger c_{n,2})\right.\\
  &+\frac{v+\delta v(t)}{2}c_{n,1}^\dagger c_{n,2}+h.c.\\
  &\left.+\frac{v-\delta v(t)}{2}c_{n,2}^\dagger c_{n+1,1}+h.c.\right]\numberthis.
\end{align*}
\begin{figure}[t]
  \subfloat[]{\label{fig:rm-structure}\includegraphics[width=0.48\textwidth]{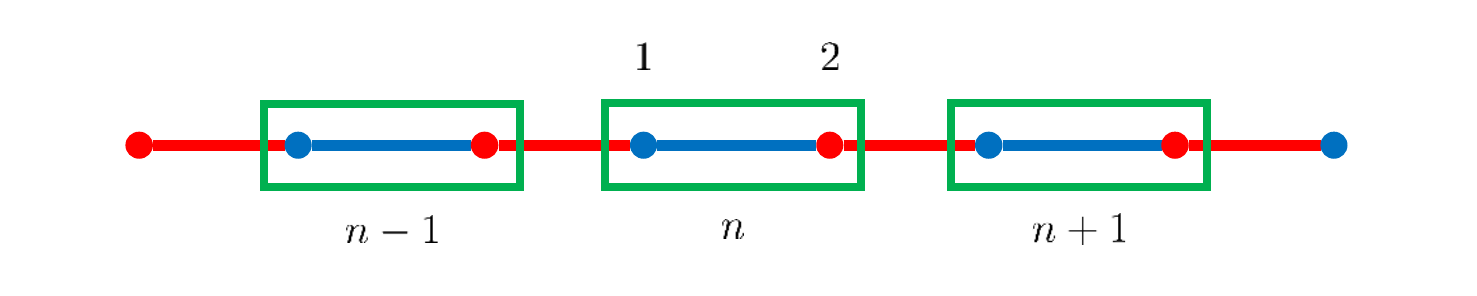}}

  \subfloat[]{\label{fig:rm-trace}\includegraphics[width=0.3\textwidth]{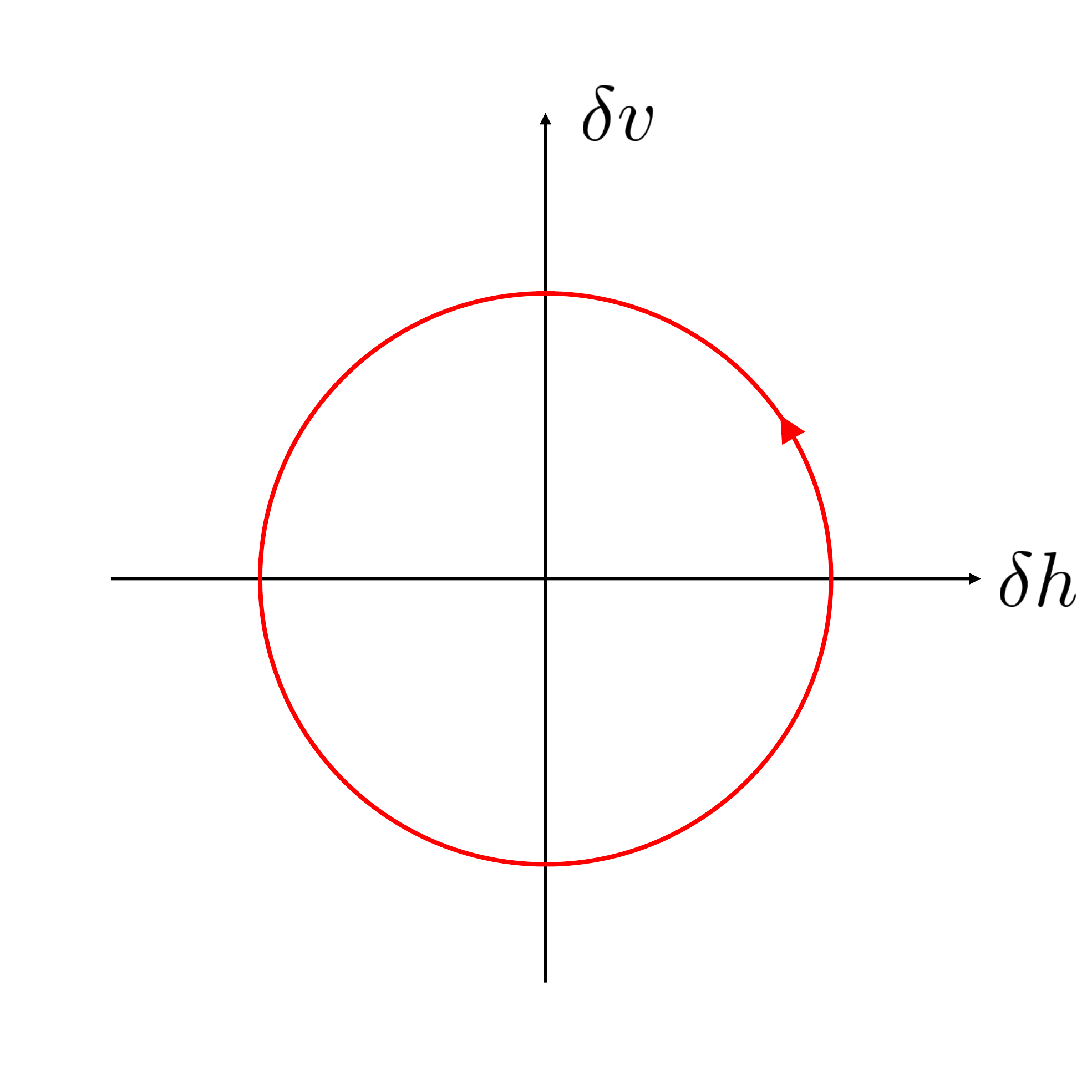}}
  \caption{\ref{sub@fig:rm-structure} Visualized structure of the Rice-Mele model. Green rectangle shows the unit cell which contains two sublattices with staggered potential $\pm \delta h$, marked by blue and red dots. Blue and red lines connecting two sites represent the alternating hopping $v\pm \delta v$.  \ref{sub@fig:rm-trace} Trace of the parameters $(\delta h, \delta v)$ when $\delta h(t)=\cos(2\pi t/T)$ and $\delta v(t)=\sin(2\pi t/T)$. In a period, the trace of parameters $(\delta h, \delta v)$ winds origin once along the anti-clockwise direction.}
  \label{fig:rm-details}
\end{figure}

Here, we modulate a staggered potential $\delta h(t)$ and an alternating hopping $\delta v(t)$ as the time $t$ changes, while the constant hopping $v$ never changes. The creation and annihilation operators on $n$-th unit cell, with sublattice $1$ and $2$, has been written as $c_{n,1/2}^\dagger, c_{n,1/2}$, respectively. Visualized structure of the Rice-Mele model is shown in Fig.\ref{fig:rm-structure}.

In the case when $|\delta h|,|\delta v|<|v|$ and at half-filling, as the path of the parameter $(\delta h, \delta v)(t)$ winds around the origin of the parameter space, this model pumps charges with time period. For example, when $\delta h (t)$ and $\delta v (t)$ are given as,
\begin{equation}\label{eqn:parameter-winding}
  \delta h(t)=\cos\left(\frac{2\pi t}{T}\right),\quad \delta v(t)=\sin\left(\frac{2\pi t}{T}\right),
\end{equation}
the charge pumping occurs in a period $T$, since the parameter set $(\delta h,\delta v)$ winds around the origin as in Fig.\ref{fig:rm-trace}.

The number of charge pumped by the Rice-Mele model coincides with the winding number of the parameter space, or equivalently, the Chern number of the Hamiltonian,
\begin{equation}
  C=\frac{1}{2\pi}\sum_{n} \int_0^T dt \int dk F_{t,k}^n.
\end{equation}
Here, the summation $\sum_n$ runs around the eigenstates below the energy gap, and $F_{t,k}^n$ represents the Berry curvature of $n$-th eigenstate, at time $t$ and momentum $k$. Because the Chern number is a topological property of the system, this shows the reason why the amount of charge is quantized.

\section{Localization near equipotential line}\label{appendix:stark-localization}
The relation between strength of frequency and the localization on equipotential line is a key point of the research. In this paragraph we show that the localization indeed occurs when the frequency is high enough.

For the parameter of localization, we choose the variation of the distance from equipotential line $\Delta x^2$. Specifically, we choose an equipotential line and define $x$ as an operator measuring the distance from the equipotential line to each cite. By calculating variance $\Delta x^2 = \langle x^2\rangle - \langle x\rangle^2$ of the state, we get the dispersion of the state around the equipotential line. Notice that the variance $\Delta x^2$ does not depend on the position of the equipotential line we take.

\begin{figure}[t]
  \includegraphics[width=0.45\textwidth]{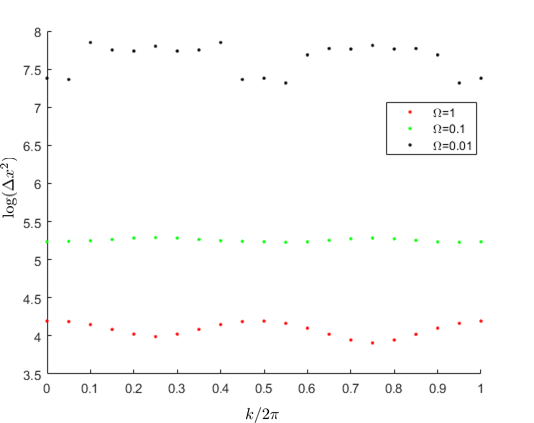}
  \caption{The log value of the variance of distance from the equipotential line  $\Delta x^2=\langle x^2\rangle-\langle x\rangle^2$  on each $k$. As the frequency increases from $\Omega=0.01$ (black dots) to $\Omega=0.1$ (green dots) and $\Omega=1$ (red dots), the variance of distance decreases. The data has been plotted on $\Delta=3, \delta(k)=\sin k, B_z(k)=0.05\cos k$, and $p=16, q=17$.}
  \label{fig:localization}
\end{figure}

As we can see in Fig.\ref{fig:localization}, the variation $\Delta x$ becomes smaller as the frequency becomes higher. This indicates that the state has a contribution from more sites near equipotential line. This corresponds to the fact that the localization on the equipotential line occurs due to the quasi-electric field, whose strength depends on the frequency. As frequency increases, the strength of quasi-electric field on the Floquet lattice also increases, which makes the Stark effect even stronger.

\bibliography{biblio}


\end{document}